\documentclass[11pt]{article}

\usepackage[english]{babel}
\usepackage[utf8]{inputenc}

\usepackage{amssymb,dsfont,amsmath,amsfonts,hyperref,mdframed,mathrsfs,stmaryrd,bbold}
\usepackage[enableskew]{youngtab}
\usepackage[active]{srcltx}
\usepackage{slashed}
\usepackage{color}
\usepackage{graphicx}

\advance \textheight by \topskip
\DeclareMathAlphabet{\mathpzc}{OT1}{pzc}{m}{it}
\usepackage{xcolor}

\usepackage[normalem]{ulem}
\usepackage{graphicx}
\usepackage{soul}
\usepackage{cite}

\usepackage{mathtools}

\usepackage{array}
\usepackage{cancel}

\definecolor{mygreen}{rgb}{0.1, 0.7, 0.2}

\def\be{\begin{equation}}
\def\ee{\end{equation}}
\def\bea{\begin{eqnarray}}
\def\eea{\end{eqnarray}}

\def\a{\alpha}
\def\b{\beta}
\def\g{\gamma}

\def\eg{\textit{e.g.} }

\def \half{spin--$\tfrac12$ }
\def \tralf{spin--$\tfrac32$ }

\newcommand \q {\rho}
\newcommand \p {\eta}

\def\spartial {\slashed{\partial}}

\def\vspin{\Psi}

\def\snabla{\slashed{\nabla}}

\def\mcC{\mathcal{C}}

\title{ {Gauge and time-reparametrization invariant spin-half fields}}

\author{Mauricio Valenzuela\footnote{Email: mauricio.valenzuela at uss.cl}\\[12pt]
\textit{Facultad de Ingenier\'ia, Arquitectura y Dise\~no, Universidad San Sebasti\'an, Valdivia, Chile}\\
and \\
\textit{Centro de Estudios Cient\'{\i}ficos (CECs), Arturo Prat 514, Valdivia, Chile} 
}

\begin{document}

\maketitle

\begin{abstract}

{We present a fermion model characterized by an anticommuting-parameter shift symmetry. The Hamiltonian formulation exhibits a combination of first-class and second-class constraints. We derive the well-known Dirac equation by fixing the gauge in a covariant manner, enabling the fields to propagate accordingly. Notably, the model inherently possesses invariance under reparametrizations of time. Consequently, the Hamiltonian vanishes, setting it apart from the conventional framework of Dirac's theory. 
Furthermore, we establish a correspondence between these particles and the zero energy modes of the massless Rarita-Schwinger system, bringing forth the intriguing implication that they may describe a supergravity ground state.}
%
%

 \end{abstract} 

\tableofcontents

\section{Introduction}

Particle physics models currently employed describe fermionic matter, encompassing leptons and quarks, as spin-half representations conforming to the Lorentz group. These representations adhere to the variational principle derived from the first-order Dirac Lagrangian. Within the framework of fiber bundles, the (non-)abelian gauge interactions are formulated, wherein the Dirac fields assume the role of local sections that transform as modules of gauge symmetry groups. Concurrently, the force carrier potentials manifest as (gauge) connections.

This article introduces an alternative characterization of spin-half fields wherein they transform as gauge connections associated with a local fermion-parametric shift symmetry, akin to the Volkov-Akulov \cite{Volkov:1973ix} or the massless Rarita-Schwinger model employed in supergravity \cite{fnf,dz,Ducrocq:2021vkh}.

{The model exhibits inherent time-reparametrization invariance, without extending the phase space, resulting in the vanishing of the energy of the spin-half field, in contrast to standard Dirac particles. This model generalizes the pseudoclassical system \cite{Valenzuela:2022zic} to higher dimensions, and we present its derivation as a truncation of the massless Rarita-Schwinger system.}

{The presence of time-reparametrization invariance raises intriguing questions about the nature of time itself \cite{Isham:1991mm}. It suggests that the concept of time, as we perceive it in our macroscopic world, may be an emergent phenomenon from a more fundamental description. Time-reparametrization invariance in theories such as quantum gravity motivates the pursuit of deeper insights into the nature of time and its role in the fundamental laws of physics. The system presented in this paper is motivated by such inquiries.}

{Moreover, we will demonstrate that the model under consideration describes the zero energy modes of the massless Rarita-Schwinger connection. This intriguing finding suggests that particles of this nature may be present in the supergravity ground state, aligning with recent findings \cite{Valenzuela:2022gbk,Valenzuela:2023ips}.}

The organization of this paper is as follows: In Section \ref{sec:model}, we describe the Lagrangian model. The Hamiltonian analysis is conducted in Section \ref{sec:Hamilton}. In Section \ref{sec:DC}, we demonstrate that this serves as a counterexample to Dirac's conjecture. Section \ref{sec:RS} presents the connection with the massless Rarita-Schwinger system. Finally, our conclusions are presented in Section \ref{sec:conc}.

\section{The model}\label{sec:model}

Consider the action principle,
\be\label{S}
S=-i\int d^Dx\, \left(\bar{\zeta} \g^0 \dot{\psi}+ \bar\zeta \snabla \theta \right) \,,
\ee
as a model for fields $\theta(t,x),\,\psi(t,x),\,\zeta(t,x)$, equivalent to $k=2^{[D/2]}$ component  spinors in $D=d+1$ spacetime dimensions. Here $(\g^\mu)^\a{}_\b$ are Dirac matrices, $\snabla:= \g^i\, \partial_i$, $i=1,\cdots,d$ and $\bar\psi_\a=(\psi^\b)^* C_{\b\a}$, where $C$ is the conjugation matrix. 

Complex spinors can be handled by adding the complex conjugate to \eqref{S}. However, we shall consider Majorana spinors for simplicity since the reality conditions will not change the properties to be studied here. Hence we shall assume that the Majorana spinors are ``really real"  ($\psi^*=\psi$), in $D=2,3,4$ Mod $8$ dimensions, and the conjugation matrix, $C_{\a\b}=-C_{\b\a}=-C^{\a\b}$, lowers and rises spinor indices according to the north-west/south-east convention,  $\bar\psi_\a=\psi^\b C_{\b\a}$, and we choose $(\g^0)_{\a\b} = \delta_{\a\b}$.

The variation of the action yields the field equations,
\be\label{ELeq}
\g^0\dot \psi+\snabla\theta=0\,, \qquad  \snabla\zeta=0\,,\qquad \dot{\zeta}=0 \,.
\ee

The action and \eqref{ELeq}  have the gauge symmetry,
\be\label{gauge}
\delta\theta=\g^0 \dot\epsilon\,,\qquad \delta\psi=\snabla\epsilon\,,\qquad \delta\zeta=0\,,
\ee
where $\epsilon$ is a fermion spinor parameter.

The action \eqref{S} is also invariant with respect to the reparametrization of time 
\be\label{rep}
t\rightarrow t'(t)\,,\qquad \theta \rightarrow \theta'^\a=\frac{\partial t}{ \partial t'} \theta^\a\,,\qquad \zeta'=\zeta\,,\qquad \psi'=\psi.
\ee
{Hence $dt\, \theta$  is} a one-form, and $\zeta$, $\psi$ are zero-forms. It follows that the action is invariant with respect to the group diff$(\mathds{R})\times {SO}(d) $, of time-diffeomorphism and the spatial rotation group. 

Even though the system is not explicitly Lorentz-covariant, its spacetime symmetry is larger, and in the gauge,
\be\label{gf}
\theta-c\psi\approx 0\,,
\ee
where $c$ is a constant, the Lorentz covariance is recovered. Indeed,  \eqref{gf} back in \eqref{ELeq} yields the Dirac equation,
\be\label{waveq}
\spartial\psi=0\,,
\ee
where $\spartial:=c^{-1}\g^0\partial_t +\snabla$ is the Dirac operator and $c$ appears as the speed of light. 

The gauge \eqref{gf} can be reached since we can always pass to a new configuration $\theta'=\theta+\g^0 \dot\epsilon$ and $\psi'=\psi + \snabla\epsilon$ with parameter
\be
\epsilon=\tilde\spartial^{-1} (\psi-c^{-1}\theta) \,,
\ee
where $\tilde \spartial:=c^{-1}\g^0\partial_t -\snabla$, that fulfills the gauge. 

In a different gauge choice, say  
\be\label{gf'}
\theta-{c'}(t)\psi\approx 0,
\ee
we get instead
\be
 \frac{\g^0}{c'(t)} \frac{\partial \psi}{\partial t} +\snabla\psi=0\,.
\ee
which is equivalent to \eqref{waveq} in a different parametrization of time $t'$, such that,
\be
 \frac{1}{c'(t)} \frac{\partial }{\partial t} =  \frac{1}{c} \frac{\partial }{\partial t'} \,.
\ee
Thus, up to the reparametrization of the time coordinate, the gauges \eqref{gf} and \eqref{gf'} are physically equivalent. 

\section{Hamiltonian formulation}\label{sec:Hamilton}

From \eqref{S} the Lagrangian function reads,
\be\label{L}
L= -i \int d^{d}x \left(\bar{\zeta} \g^0 \dot{\psi}+ \bar\zeta \snabla \theta \right).
\ee

The Legendre transform,
\be
(\pi_\theta)_\a=\frac{\delta L}{\,\delta \dot \theta^\a}\approx 0\,,\qquad (\pi_\zeta)_\a=\frac{\delta L}{\,\delta \dot \zeta^\a}\approx 0\,,\qquad (\pi_\psi)_\a=\frac{\delta L}{\,\delta \dot \psi^\a}=   i (\bar\zeta\g^0)_{\a} \,,
\ee
produces  the \textit{primary constraints} 
\be\label{pric}
\pi_\theta\approx 0\,,\qquad \chi_1=\pi_\zeta \approx 0\,,\qquad \chi_2=\pi_\psi- i\bar\zeta\g^0 \approx 0\,,
\ee
which describes the phase space subvariety containing the physical degrees of freedom. 

The unconstrained phase space comes with the Poisson bracket,
\bea
\{f(t,x),g(t,y)\}&=& (-1)^{|f|}  \int d^{s}z \left(\frac{\delta f(t,x)}{\delta \q^a(t,z)} \frac{\delta g(t,y)}{\delta \p_a(t,z)}+ \frac{\delta f(t,x)}{\delta \p_a(t,z)}\frac{\delta g(t,y)}{\delta \q^a(t,z)} \right)\,,
\eea
where $(\q^a;\p_a)=(\theta\,,\psi,\,\zeta;\pi_\theta,\,\pi_\psi,\,\pi_\zeta)$, $a=1,2,3$, and $|f|=0,1,$ is the even, odd, Grassmann parity of the function $f$.

The constraints $\chi_I\,,I=1,2$ are second-class, 
\be
\{\chi_{I},\chi_{J }\}= \mcC_{IJ} \,,\qquad \mcC_{IJ}:= -i \left(
\begin{array}{cc}
0 & (\g^0)_{\a\b}\\
(\g^0)_{\a\b} & 0 
\end{array}\right),
\ee
where $(\g^0)_{\a\b} =- (C\g^0)_{\a\b}$, since  $\mcC_{IJ}$ is invertible,
\be\label{Ci}
\mcC^{-1\, IJ}=i \left(
\begin{array}{cc}
0&(\g^0)^{\a\b}\\
(\g^0)^{\a\b} &0
\end{array}\right)\,.
\ee
The constraint $\pi_\theta\approx 0$ is first-class since $\{\pi_\theta,\chi_I\} \approx 0$.

The canonical Hamiltonian reads, 
\bea
H_0&=& \int d^{d}x(\dot\theta ^t \pi_\theta +\dot\psi ^t \pi_\psi +\dot\zeta ^t \pi_\zeta) - L ,\\
&=& i  \int d^{d}x\, \bar\zeta\snabla\theta.
\eea
Implementing the primary constraints,
\be\label{HT}
H_T:= H_0 + \int d^{d}x(\pi_\theta \nu + \chi_I \mu^I  )\,,
\ee
with Lagrange multipliers $\nu$ and $\mu^I$, defines the \textit{total Hamiltonian}.

Thus the action principle, in Hamiltonian form, is given by,
\bea\label{Haction}
S_H&=&\int dt \left(  \int d^{d}x ( \dot\theta \pi_\theta +\dot\psi \pi_\psi +\dot\zeta \pi_\zeta) - H_T\right).
\eea
The field equations can be obtained from the action principle, and more general functions in the phase space evolve according to,
\be \label{evo}
\dot f =\partial_t f + \{f,H_T\}\,.
\ee
The  stationary conditions for the primary constraints are given by,
\bea
&\dot \pi_\theta=i\bar\zeta\overleftarrow{\snabla}\approx 0\,,&\label{sec}\\[5pt]
&\dot \chi_1=0\quad \Rightarrow \quad \mu^2=\g_0\snabla \theta \,,\qquad \dot \chi_2=0\quad \Rightarrow \quad \mu_1=0\,.&\label{eqchi}
\eea 
Equation \eqref{sec} is a secondary constraint. In \eqref{eqchi}, we obtain the solution for the second-class Lagrange multipliers---as a function of the remaining phase space variables---from their stationary conditions.

The fields evolve according to,
\bea
& \dot \psi =-\mu^2\,,\qquad \dot \zeta=-\mu^1\,,&\label{eom1} \\ 
&\dot \theta=-\nu \,.&\label{eom2}
\eea 
From \eqref{sec}, \eqref{eqchi} and \eqref{eom1}, we reproduce the Euler-Lagrange equations \eqref{ELeq}. 

\subsection{Gauge fixing}\label{sec:red}

The gauge ambiguities are reflected in $\nu$ and $\theta$.  With the gauge choice \eqref{gf}, $\theta=c\psi$ ceases to be independent, and we recover the Dirac equation \eqref{waveq} from \eqref{eqchi} and \eqref{eom1}.

The stationary condition of \eqref{gf}, $\dot\theta-c\dot\psi\approx0$ fixes the value $\nu=c\mu^2$, together with \eqref{eqchi} we obtain,
\be\label{nu}
\nu=c^2\g_0\snabla\psi,
\ee
and there are no arbitrary free fields left in the system.

Alternatively, restricting the system to the surface $\chi_I=0$ \eqref{pric}, the variables $\pi_\zeta\approx 0$ can be removed, and $\zeta$ can be replaced by $\pi_\psi$. Hence, the Dirac bracket $\{f,g\}_D:= \{f,g\} - \{f,\chi_I\} \,\mcC^{-1\,IJ}\{\chi_J,g\}$ reduces to
\be \label{DB}
\{f,g\}_D= ( - 1)^f \left(\frac{\partial f}{\partial \theta} \frac{\partial g}{\partial \pi_\theta}+ \frac{\partial f}{\partial \pi_\theta}\frac{\partial g}{\partial \theta} +\frac{\partial f}{\partial \psi} \frac{\partial g}{\partial \pi_\psi}+\frac{\partial f}{\partial \pi_\psi}\frac{\partial g}{\partial \psi} 
\right)\,,
\ee
on functions of $f(\theta,\pi_\theta;\psi,\pi_\psi)$. Thus the relevant canonical relations are, 
 \be\label{alg}
\{\theta^\a ,(\pi_\theta)_\b\}_D=-\delta_\b^\a\,,\qquad  \{\psi^\a,(\pi_\psi)_\b\}_D={-\delta_\b^\a\,.}
\ee
The reduced Hamiltonian, in the surface of strongly vanishing $\chi_I$, is given by,
\be\label{rH}
H_R:=\int d^{d}x\,  \left( \pi_\psi \g_0\snabla \theta +\pi_\theta \nu\,\right).
\ee
From $\pi_\theta\approx 0$, $\dot\pi_\theta$ implies $ \pi_\psi\overleftarrow{\snabla}\approx 0$ and we obtain the field equations,
\be\label{HReq}
\g^0\dot \psi+\snabla\theta\approx 0\,, \qquad  \pi_\psi \approx 0 \,, \qquad  \dot\theta = -\nu \,.
\ee
With the gauge fixing \eqref{gf} we solve for $\nu$ as in \eqref{nu}, and the gauge condition yields from \eqref{HReq} the Dirac equation \eqref{waveq}.

The constraints $\pi_\theta=0$ and the gauge \eqref{gf} set strongly are algebraic constraints that project the phase space to the subspace spanned by coordinates $(\psi,\pi_\psi)$. In this subspace, the Dirac bracket reduces to,
\be\label{rDB}
\{f,g\}_R= (-1)^f \left( \frac{\partial f}{\partial \psi} \frac{\partial g}{\partial \pi_\psi}+\frac{\partial f}{\partial \pi_\psi}\frac{\partial g}{\partial \psi} 
\right)\,.
\ee

The gauge-fixed Hamiltonian reads,  
\be\label{Hclas}
H_{fix}=c \int d^{d}x\, \pi_\psi\g_0 \snabla \psi\,,
\ee
and now the fields evolve according to,
\be\label{eomeff}
\dot f= c (-1)^{|f|} \left( \frac{\partial f}{\partial \psi} \cdot \g_0 \snabla \psi- \frac{\partial f}{\partial \pi_\psi} \cdot (\pi_\psi \overleftarrow{\snabla}) \g_0
\right)\,.
\ee
Considering the secondary constraint $\pi_\psi(t,\vec x)\approx 0${, the Hamiltonian vanishes. The evolution of the system is nontrivial and given by},
\be\label{eomeff2}
\dot f= c  (-1)^{|f|} \frac{\partial f}{\partial \psi} \cdot \g_0 \snabla \psi \,,\qquad  \pi_\psi\approx 0\,,
\ee
on functions $f(\psi,\pi_\psi)$, and we set $\pi_\psi=0$ in the end of the computation.

Note that in the Hamiltonian formulation of the standard Dirac theory, the second-class constraints, $\pi_\psi \approx \bar\psi$, remove the conjugate momenta differently{, and the Hamiltonian does not vanish}. However, the evolution of the system is still given by the first equation in \eqref{eomeff2}.

\subsection{The secondary first-class constraint as an initial condition}\label{sec:initc}

Equation \eqref{eomeff2} is equivalent to \eqref{eomeff} with initial condition $\pi_\psi(t=0,\vec x)=0$, 
\be\label{eomeff3}
\dot f= c (-1)^{|f|} \left( \frac{\partial f}{\partial \psi} \cdot \g_0 \snabla \psi- \frac{\partial f}{\partial \pi_\psi} \cdot (\pi_\psi \overleftarrow{\snabla}) \g_0
\right)\,,\qquad \pi_\psi(0,\vec x)=0\,,
\ee
since
\be\label{pieq}
\dot \pi_\psi= c \pi_\psi \overleftarrow{\snabla}\g_0\,, \qquad  \pi_\psi(0,\vec x)=0\,,
\ee
implies
\be\label{dn}
\frac{d^n\pi_\psi}{dt^n}(0,\vec x)=0\,,
\ee
for all order derivatives at the initial Cauchy surface. Hence, from its Taylor expansion, around $t=0$, $\pi_\psi(t,\vec x)$ vanishes for all $t$.  This means, 
\be\label{pipsi}
\pi_\psi(t,\vec x)=0\,,
\ee
is the only solution to the field equations \eqref{eomeff3} with initial condition $\pi_\psi(0,\vec x)=0$. 
 
\section{Another exception to Dirac's conjecture}\label{sec:DC}

Dirac conjectured in  \cite{DiracQM} that ``it may be that all the first-class secondary constraints should be included among the transformations which don't change the physical state". Then, regarding primary and secondary first-class constraints in the same footing, Dirac proposed the \textit{extended Hamiltonian}, which includes primary and secondary first-class constraints, as the system's time translation generator. {The inclusion of secondary first-class constraints introduces undetermined velocities, equivalent to the Lagrange multipliers associated with these constraints. As a consequence, the corresponding fields become arbitrary functions of time and are thus excluded from the physical spectrum.}

{In Dirac's treatment of Hamiltonian systems, the degrees of freedom of the model presented here would all become unphysical. As his approach is so widely adopted, it is essential stressing that this model is not an example of his conjecture.}
 
Noticing that the secondary constraint \eqref{sec} $\snabla\zeta\approx 0$ implies $\zeta\approx 0$ on the space of non-constant functions of $\vec x$---where $\snabla$ is invertible---the constraints \eqref{pric}-\eqref{sec}, $(\pi_\theta,\chi_I,\zeta,)\approx 0$ can be rearranged as $(\pi_\theta,\pi_\zeta,\pi_\psi,\zeta)\approx 0$ and now $(\pi_\zeta,\zeta)\approx 0$ are second-class while $(\pi_\theta,\pi_\psi)\approx 0$ are first-class.  In Dirac's extended dynamics, the {conjugate variables} associated with the first-class constraints $\pi_\psi\approx 0$ and $\pi_\theta \approx 0$,  { respectively $\psi$ and $\theta$, would both become arbitrary functions of time, equivalent to gauge redundancies.}

However, the secondary first-class constraint  $\pi_\psi\approx 0$ does not generate an independent gauge symmetry. The transformation $\delta\lambda = \epsilon$---{affecting the field equations generated by the total Hamiltonian} \eqref{eqchi} and \eqref{eom1}---would require $\dot\epsilon=0$ to produce a symmetry, then $\epsilon$ is not arbitrary. 

The true gauge symmetry, \eqref{gauge}, is generated by the Castellani chain \cite{Castellani:1981us},
\be\label{G1}
G(\epsilon)=\int d^dx \left(\pi_\theta  \g^0 \dot\epsilon +\pi_\lambda \snabla \epsilon\right).
\ee
Hence the number of {gauge orbits, of completely indeterminate phase space directions,} is equivalent to the number of components in $\epsilon$. This allowed us to fix $\theta$ in terms of $\psi$ in \eqref{gf} and the system became deterministic. Consequently, imposing a secondary gauge condition would be unjustified. {Evidently, the presence or not of the Dirac field $\psi$ represents different physicals state of the system, contradicting Dirac's conjecture.}

 Though counterexamples of the Dirac conjecture are well known \cite{Cawley:1979oou,Frenkel:1980nt,Gotay,Henneaux:1992ig}, it is argued (\eg in \cite{Henneaux:1992ig}) that it is better to assume Dirac's conjecture as a general principle, to prevent issues concerning the ill-definition of the Poisson bracket on phase space subspaces where some coordinates have no conjugate momenta, implying an obstruction for quantization. 
 
{In the approach followed in section \ref{sec:initc} this situation was avoided by setting secondary first-class constraints as initial conditions \eqref{pieq}.}  {It is worth noticing that the insight that a secondary first-class constraint can be viewed as a `mere initial condition' is pointed out in Weinberg's book on Quantum Field Theory \cite{wein-qft}, as referring to the electromagnetic Gauss law. Reference \cite{Valenzuela:2023ips} extends this argument to encompass secondary first-class constraints in general, demonstrating its crucial role in the canonical analysis and quantization of Dirac's conjecture counterexamples. This approach has been employed to quantize the counterexamples to Dirac's conjecture found in \cite{Valenzuela:2023ips}, including Cawley's system \cite{Cawley:1979oou}, and the one-dimensional analog \cite{Valenzuela:2022zic} of the model under consideration. }

\section{Relation to the massless Rarita-Schwinger Lagrangian}\label{sec:RS} 

{Here we demonstrate that the characterization of \half particles presented here describes the lowest energy state of the massless Rarita-Schwinger system,}
\be\label{RSL}
\mathcal{L}:=-\frac i 2  \bar\vspin_\mu  \gamma{}^{\mu\nu\lambda}\partial_\nu \vspin_\lambda\,,
\ee
{as referred to in the context of supergravity \cite{fnf,dz,van}\footnote{Note however that the \tralf theory proposed initially by Rarita and Schwinger \cite{r-s} differs from \eqref{RSL} (for further details see \cite{Weinberg} and \cite{Valenzuela:2022gbk,Valenzuela:2023aoa}).}
Here $\psi_\mu^\a $ is a vector-spinor.}

Splitting space and time directions we obtain,
\be\label{RSL3}
\mathcal{L}= -i\bar\vspin_0 \gamma^{0ij} \partial_i \vspin_j +\frac i 2 \bar \vspin_i \gamma^{0ij} \dot{ \vspin}_j -\frac i 2 \bar \vspin_i \gamma^{ijk} \partial_j \vspin_k\,.
\ee
We can use the projector operators introduced in \cite{Valenzuela:2022gbk},
\begin{align}\label{proj}
(P^N{})_{ij}:= \frac{1}{D-2}N_i N_j\,,\qquad (P^L{})_{ij}:&=  L_i L_j , \qquad P^{T}=\mathds{1}- {P}^N- P^L\,,
\end{align}
where $N_i:=\g_i-L_i$ and $L_i:=\snabla^{-1} \partial_i$,  satisfying the identities $N_i\, N^i =D-2$, $L_i \,L^i = 1$, $N_i\, L^i =0$,
to further decompose the spatial vector-spinor $\Psi_i$ as 
\be \label{psidec1}
\Psi_i = \xi_i + N_i \zeta + L_i \lambda\,, 
\ee
where
\be \xi_i= P^{T} {}_i{}^j \Psi_j \,,\qquad \zeta = \frac{1}{D-2}N^i \Psi_i\,,\qquad \lambda = L^i\Psi_i\,,
\ee
carry representations of spin $\frac32$,  $\frac12$ and $\frac12$ of the spatial rotation group. Thus we obtain
\be\label{Ldec}
\mathcal{L}= i \int d^{D-1}x \; \left[ -(D-2) \bar\Psi_0 \g^0  \slashed{\nabla}\zeta + (D-2) \bar\lambda\g^0\dot\zeta-\frac12\bar\xi^{i} \slashed\partial \xi_i +\frac{(D-2)(D-3)} 2 \bar\zeta \slashed\partial \zeta  \right] \,.
\ee
The field equations read,
\be\label{dxi}
\slashed\partial \xi_i=0\,,
\ee
where $\g^i\xi_i=\partial^i\xi_i=0$, and
\bea\label{eomm}
\snabla\zeta=0\,,  \qquad \dot\zeta=0\,,\qquad \g^0 \dot\lambda +\slashed\nabla \g^0\Psi_0=0\,.
\eea
The system \eqref{eomm} reduces to \eqref{ELeq} with $\lambda=\psi$ and $\g^0\Psi_0=\theta$. Hence, and from the results of section \ref{sec:Hamilton}, \eqref{eomm} describes a \half\!\! particle.

The field $\xi_i$ is the \tralf (double-transverse) mode found in reference \cite{Deser:1977ur}. However, assuming the validness of the Dirac conjecture, the \half sector \eqref{eomm} was missed in the earliest literature \cite{Senjanovic:1977vr,Pilati:1977ht,Deser:1977ur,Fradkin:1977wv}, as it was understood in \cite{Valenzuela:2022gbk,Valenzuela:2023aoa}. 

The model  \eqref{L}  can be obtained from the Rarita-Schwinger Lagrangian truncating \eqref{Ldec} to the sector $\xi_i=0$,
\be\label{Ldec'}
\mathcal{L}\vert_{\xi_i=0}= i \int d^{D-1}x \; \left[ -(D-2) \bar\Psi_0 \g^0  \slashed{\nabla}\zeta + (D-2) \bar\lambda\g^0\dot\zeta+\frac{(D-2)(D-3)} 2 \bar\zeta \slashed\partial \zeta  \right] \,.
\ee 
In contrast with \eqref{L},  \eqref{Ldec'}  is not time-reparametrization invariant, from the presence of $ \bar\zeta \slashed\partial \zeta$. However, both Lagrangians produce the same on-shell physics. Indeed, removing the last term in \eqref{Ldec'} does not affect the field equations; up to boundary terms and a dimensionless constant, we recover \eqref{L}.

{The total Hamiltonian of the Rarita-Schwinger action can be decomposed into \tralf and \half decoupled components \cite{Valenzuela:2022gbk,Valenzuela:2023aoa}, 
\be
H_{RS}=H_{3/2} + H_{1/2}\,,
\ee
where $H_{3/2}$ is the \tralf particle Hamiltonian, $\xi_i$, and $H_{1/2}$ is equivalent to the Hamiltonian \eqref{HT}, of \half particles. The Hamiltonian $H_{3/2}$ does not vanish for non-trivial configurations $\xi_i$, while $H_{1/2}$ vanishes, owing to the presence of the constraints $\pi\approx0$, even for non-trivial \half configurations, as observed in section \ref{sec:Hamilton}.
Thus, the lowest energy state of the RS system consists of the \half particles described in this paper.}

{Since \eqref{RSL} describes the fermion sector of simple supergravity (up to the metric's determinant factor), an intriguing argument can be made that the particle system discussed thus far might serve as an effective description of the fermion sector in the low-energy regime of supergravity.}

\section{Conclusions and final remarks}\label{sec:conc}

{We have introduced a spin-half particle model characterized by a fermion-parameter gauge symmetry, time-reparametrization invariance, and a vanishing Hamiltonian.}

{It is worth noting that reparametrization-invariant systems can arise in two different ways \cite{Henneaux:1992ig}. The first approach involves achieving time reparametrization invariance by introducing new degrees of freedom to a non-invariant system, promoting the time to a canonical variable along with its corresponding (constrained) conjugate momenta. This can always be accomplished. The second approach is when the system is already given in an invariant form, as is the gravity case and as demonstrated in our model.}

{The presence of time-reparametrization invariance leads to the vanishing of the Hamiltonian. Consequently, the energy of the particles described in this model is lower than that of standard Dirac fields.  Furthermore, we have shown that these particles can be identified with the lowest energy states of the massless Rarita-Schwinger system, which suggest that they can arise in supergravity on the same energy level as gravitational degrees of freedom. This conclusion aligns with previous findings that supergravity models exhibit truncations to spin-half fields in gravitational backgrounds \cite{Alvarez:2011gd,Alvarez:2020qmy,Alvarez:2021qbu,Valenzuela:2022gbk,Valenzuela:2023aoa}, also known as \textit{unconventional supersymmetry} models \cite{Alvarez:2020izs,Alvarez:2021zhh}.}

\section*{Acknowledgements}

We thank discussions with J. Zanelli. This work was partially funded by grant FONDECYT 1220862. 
We want to thank also the anonymous referees for their valuable feedback which significantly contributed to improving the quality and clarity of this manuscript.

\bibliographystyle{unsrt}


\end{document}